# Graphical Abstract

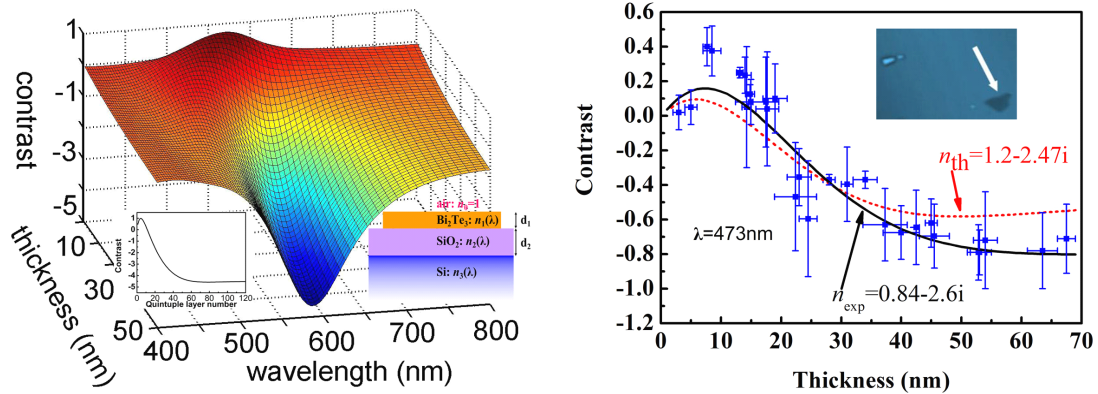

Topological insulator (TI) is a new class of quantum materials which band structure exhibits a nontrivial $Z_2$ topology. Such materials possess a single Dirac cone located inside the bulk bandgap and therefore the potential applications in the spintronics. Mechanical exfoliation is a convenient method to prepare the mesoscale TI samples inheriting the bulk crystalline structures, where a critical issue lies to locate the TI flakes of interests. Optical reading has been demonstrated in the study of graphene. Here we extend the approach to the study of mesoscale TI flakes. The left figure shows the calculated contrast of $Bi_2Te_3$ flakes on a $SiO_2$-capped silicon wafer. It is confirmed by the experiment of the optical contrasts dependent on the thickness as shown in the right right. One may see the contrast inversion in its inset. Please see the detailed discussion in the text.

# Visualizing topological insulating $Bi_2Te_3$ quintuple layers on $SiO_2$-capped Si substrates and its contrast optimization

Zhaoguo Li, Yuyuan Qin, Yuewen Mu, Taishi Chen, Changhui Xu, Longbing He, Jianguo Wan, Fengqi Song[*], Min Han, Guanghou Wang

Department of Physics and National Lab of Solid State Microstructures, Nanjing University, 210093, Nanjing, P. R. China

Abstract: Thin $Bi_2Te_3$ flakes, with as few as 3 quintuple layers, are optically visualized on the $SiO_2$-capped Si substrates. Their optical contrasts vary with the illumination wavelength, flake thickness and capping layers. The maximum contrast appears at the optimized light with the 570nm wavelength. The contrast turns reversed when the flake is reduced to less than 20 quintuple layers. A calculation based on the Fresnel law describes the above observation with the constructions of the layer number-wave length-contrast three-dimensional (3D) diagram and the cap thickness-wavelength-contrast 3D diagram, applicative in the current studies of topological insulating flakes.

[*] Corresponding author: songfengqi@nju.edu.cn

$Bi_2Te_3$, a narrow band-gap semiconductor, has been featured as a vital component for the thermoelectric industry since the bulk $Bi_2Te_3$-based materials possess the highest thermoelectric figure of merit (ZT~1.14) at room temperature[1-5]. It has recently gained even more attention from the physics community after demonstrated to be a 3D topological insulator (TI)[6-8], which is a new state of quantum matter with a prominent bulk energy gap and a conductive surface state simultaneously[9-11]. Such a surface state even presents a Dirac cone in its electronic structure[12-13]. Due to the entanglement between the carriers in the top and the bottom layer, the TI behavior of $Bi_2Te_3$ exhibits the obvious dependence on the number of the quintuple layers[6,14], which thus expects precise layer control during the sample preparation or relocating the nanosheets. The $Bi_2Te_3$ crystals is of the rhombohedral structure, where five atomic layers ($Te^{(1)}$-Bi-$Te^{(2)}$-Bi-$Te^{(1)}$) form a quintuple unit with thickness ~1nm. Between two quintuple layers (QLs), the spacing is larger and the weak van der Waals interaction ($Te^{(1)}$-$Te^{(1)}$) is dominant[3,15]. This creates a possibility for producing quasi two-dimensional atomic quintuple layers by simple mechanical exfoliations. Gently rubbing the layered materials against another stiff surface may slice the crystal for few-layer materials as done in the graphene fabrication[16]. Such exfoliation-based preparation remains critical since such slices have demonstrated good crystalline structures and reliable carrier behaviors despite the fact that chemical bath deposition[17], solvothermalization[18] and various evaporation techniques[5] have formed the $Bi_2Te_3$ crystals because these techniques normally produce polycrystalline films for thermoelectric applications, far from the current TI studies[9].

However, the exfoliated thin $Bi_2Te_3$ layers are often mixed with the flakes of variable layer numbers[15]. Therefore an important issue for the exfoliated sample is to locate the TI flakes of interests, on which modern microscopic techniques, including atomic-force microscopy, scanning tunneling microscopy, and electron microscopy, lose their efficiencies due to their extremely low throughput[19-20]. In the research of graphene, an optically locating of the few-layer sheets on top of an oxidized Si wafer is a critical step[21-22]. The thin flakes are transparent to the reflected light, which contributes an additional optical path and changes their interference color contrast with respect to an empty wafer. The thickness of the thin graphite flakes can therefore be distinguished by reading their color contrast. Even the few-layer graphene can be visible in this way [21-22]. The optically-assisted locating is also applicable and a critical step to isolate the TI sheets of interests in the exfoliated $Bi_2Te_3$ quintuple layers[15], while no extensive study is reported. Furthermore, such study of locating the sub-wavelength-thick sheets by visible illumination microscopy is still of good interests in elementary optics. Here we report the visualization of the exfoliated $Bi_2Te_3$ flakes, with as few as 3 layers, by using the 300nm-thick silica capped silicon substrate and visible illumination. A calculation based on the Fresnel optics is carried out with the construction of the three-dimensional layer number-wave length-contrast diagram. The predictions on the optimized incident wavelength and the contrast reversion are largely confirmed by our experiment.

The $Bi_2Te_3$ thin sheets were prepared by a mechanical exfoliation procedure similar to the graphene preparation[16]. Such sheets have been regarded as a TI candidate recently[3,14]. A scotch tape with the crystalline $Bi_2Te_3$ powder (purchased from Alfa Aesar, vacuum deposition grade, 99.999% (metals basis)) was folded many times and then pressed onto the silica-capped wafer. The samples were then imaged by a NTEGRA Probe NanoLaboratory system (NT-MDT, Co.), where the atomic force microscopy (AFM) unit, optical imaging and confocal Raman scattering unit are combined. Monochromatic illumination is implemented by narrow-band filters (with the half widths of 10nm) as well as a few low-power lasers. **Figure 1** shows a $Bi_2Te_3$ thin sheet with 10 quintuple layers, which thickness was measured by AFM (see Fig. 1a). Its composition is confirmed by the Raman scattering[23]. We can clearly see the sheet on the silica-capped wafer upon the illumination of a blue laser (Fig.1b), while we find it becomes invisible if using an empty wafer. We also see other sheets with as few as 3 quintuple layers. This demonstrates the applicability of the visualization technique[22].

Such visualization can be described by the optical contrast driven by the Fresnel's Law[21-22,24]. We carried out the calculation, where the contrast is defined as

$$C(\lambda) = \frac{R_{SiO_2/Si}(\lambda) - R_{Bi_2Te_3}(\lambda)}{R_{SiO_2/Si}(\lambda)} \quad (1).$$

The function $R(\lambda)$ ( $R(\lambda) = r^*(\lambda)r(\lambda)$ ) is the reflected optical intensities with the light (wavelength of $\lambda$). $d_1$ and $d_2$ as defined in the right inset of **Figure 2a**. The function

$$r(\lambda) = \frac{r_1 e^{i(\beta_1+\beta_2)} + r_2 e^{-i(\beta_1-\beta_2)} + r_3 e^{-i(\beta_1+\beta_2)} + r_1 r_2 r_3 e^{i(\beta_1-\beta_2)}}{e^{i(\beta_1+\beta_2)} + r_1 r_2 e^{-i(\beta_1-\beta_2)} + r_1 r_3 e^{-i(\beta_1+\beta_2)} + r_2 r_3 e^{i(\beta_1-\beta_2)}} \quad (2).$$

The parameters $r_1 = \frac{n_0 - n_1}{n_0 + n_1}$, $r_2 = \frac{n_1 - n_2}{n_1 + n_2}$, $r_3 = \frac{n_2 - n_3}{n_2 + n_3}$, $\beta_1 = \frac{2\pi n_1 d_1}{\lambda}$ and

$\beta_2 = \frac{2\pi n_2 d_2}{\lambda}$ are calculated if the complex refractive indexes $n_0$, $n_1$, $n_2$, $n_3$ (which correspond air, $Bi_2Te_3$, $SiO_2$, Si, respectively and $n_1$, $n_2$, $n_3$ depend on $\lambda$.) are given as in the references[21-22,25]. The detailed complex refractive indexes are found to vary with different samples with different crystalline conditions, the compositions and even the substrates, where the experimental values need further interpolation[5,26]. Here the complex refractive indexes of the $Bi_2Te_3$ crystal are therefore generated using the Cambridge Sequential Total Energy Package[27]. The calculated refractive indexes are largely the same as those adapted from the experiments on the thermal evaporated samples except that the refractive index exhibits 50 percent larger fluctuation than that of the experimental values (1-2)[5,26]. As a manifestation, we change the slope of the wavelength–refractive index curve from 1/100 nm to 0, the optimal wavelength with the maximized contrast is found to shift only 5 nm. Hence, the employment of the simulated refraction indices introduced reasonable errors in the calculated contrast functions, on which we can make comparisons to the experiments. The other refractive indices (silica, silicon) were obtained from the reference book[25]. Fig. 2a and 2b show the calculated diagram of the wavelength-thickness-contrast relation and the diagram of the optical contrast plotted against the thickness of the capping layer and the incident wavelength respectively. The optical contrast of the flake obviously changes with both the light wavelength and flake thickness. We can see a common optimized wavelength in Fig. 2a. The inset left in Fig. 2a show the relation of optical contrast against the layer number with 570nm monochromatic light incidence. With the increase of the quintuple layers, the contrast increased until six-quintuples which reached the maximum contrast. Then, the contrast showed a decreased trend, and saturated when quintuple layers larger than 80. Upon the changing of the thickness of the capping layer, a few bands with the maximized contrast appear in Fig. 2b. The bands cover a wide range of cap thickness from 70-130nm and 200-350nm etc. There are effective contrasts in the light wavelength from 450nm to 640nm for our 300nm-silica-capped wafer, guaranteeing the visibility of our thin flakes. We believe even the contrast of a single quintuple layer may justify the detection by the CCD if the incidence is intense enough.

The optical contrast of the sub-wavelength-thick flakes comes from the added optical path lengths because of the difference of optical path length between $Bi_2Te_3$ and $SiO_2$ when monochromatic light incidence as the difference of refractive index of $Bi_2Te_3$ and $SiO_2$. As the refractive indexes of $Bi_2Te_3$, $SiO_2$ and Si dependent on $\lambda$ and therefore optical contrast presents a wavelength-dependent behavior. One of the critical issues is to obtain an optimized wavelength with the maximized contrast. The calculation shows the maximized optical contrasts fall in the wavelength region from 550-600nm whichever thickness of the TI flakes are imaged (Fig. 2a). We have also mentioned the optimized wavelength is predicted to be independent on the variation of the refractive indexes. Here illuminating the samples by the monochromatic light from the narrow-band filters, with the center wavelengths at 570, 524 and 470nm, the optical micrographs of a selected wafer with a few thick $Bi_2Te_3$ flakes are shown in **Figure 3a**, 3b and 3c respectively. There is a flake with the thickness of larger than 100nm as marked by the light arrow, and two thinner flakes (80 and 50nm respectively) in the left down of the graph as measured by AFM. One can see the contrast obviously increases with increasing the thickness in Fig 3a. The maximized contrasts appear at 570nm for all the shown flakes. The highlighted flake is even invisible during the illumination of 470nm (Fig 3c). The flakes are also invisible using the illumination of incident light with the wavelength larger than 600nm. An optimized illumination wavelength of 570nm that commonly provides maximized contrast is shown.

It is quite prominent that the contrast turns reversed after the critical thickness of 20nm as seen in Fig 2a. In the thinner region of more interests in the layer-dependent TI behaviors, a saturated reversed contrast is predicted at 13nm (i.e. 13 quintuple layers). The value could be a bit inaccurate because the dielectric response changes abruptly for the flake with a few quintuple layers[5]. The normal contrast increases with increasing the flake thickness in the thicker region as observed in Fig. 3. It is totally saturated at the thickness of larger than 80nm. Such contrast reversion can experimentally be observed as shown in **Figure 4a**, where a sample is illuminated by a laser at 473nm. The image is obtained by the scanning confocal mode. The few-layer TI sheets contribute much smaller contrast than the thick flakes, which is predicted to be around a fifth (Fig 2a). Therefore, more intense source is required to feed the CCD detector with a better signal/noise ratio than that with the filtered illuminations. AFM measures the thickness of less than 30nm for the flakes as marked by the white arrows in Fig 4a. The top one is 20nm thick and the below one is about 5 nm. As compared to the flakes with the thickness of more than 50nm, one may see the contrast has been reversed. After the contrast reversion, the thickness-contrast relation is no longer monotonic. Please see Fig. 1 for the images of the 10nm-thick flake, which contrast is even stronger than that of the 20nm-thick flake in Fig 4a. All the observations on the thickness-contrast relation, both the monotonic increase of normal contrast and a maximized reversed contrast, confirm the theory. We note that even the thin flakes with as few as 3 quintuple layers can be located by this approach although the contrast is extremely poor as also expected by the calculation shown in Fig. 2a. Therefore, the present approach forms a reliable technique to distinguish the few-quintuple-layer $Bi_2Te_3$ sheets by simply reading its contrast.

The applicable range and the error with the approach were assessed in order to determine the thickness by the optical reading independently. Using the 473-nm-wavlength laser, we measured dozens of samples for their optical contrast. The thickness of the flakes were determined by AFM. The experiment data were plotted in **Figure 4b**, where we also show the simulated contrast curve using the first-principles calculated refractive index $n_{th}$=1.2-2.47i (the red-dot line). The black curve in Fig. 4b was simulated using the experimentally-determined refractive index $n_{exp}$=0.84-2.6i. We can see the consistent trends of all the experimental and theoretical curves. All the curves present a minimum contrast (nearly zero) in the range of 15-20 QLs. The systematic deviation between the two calculated curves may be ascribed to the different refractive index of the subwavelength-thick flakes from that of the bulk[5]. The same source applies to the higher experimental contrasts for the few-layer flakes. An applicable range of the approach is estabilished to be 5-50 QLs, above which no resolvable difference can be measured. The flakes are no longer TIs below 5 layers[14]. In the selected range, the contrast presents a monotone decrease with the increasing QLs. The errors of such determinations are around 40 percent. The coarse approach provides a critical and effective locating step in the study of TI $Bi_2Te_3$ flakes.

In summary, we have studied visualizing the sub-wavelength-thick $Bi_2Te_3$ flakes by optical imaging the flakes on a capped Si wafer. The added optical length contributes obvious contrast even for the flake with as few as 3 quintuple layers. We obtain the layer number-wave length-contrast diagram and the wave length-cap thickness -contrast diagrams of the thin $Bi_2Te_3$ flakes supported by a silica-capped Si wafer by a calculation based on the Fresnel law. Both the optimized wavelength giving a commonly maximized contrast and the thickness-dependent contrast reversion have been demonstrated by our experiments. The present work distinguishes the few-quintuple-layer $Bi_2Te_3$ sheets by simply reading its contrast during the 570nm illumination and thus provides a critical locating step in the study of TI $Bi_2Te_3$ flakes.

We thank the National Natural Science Foundation of China (Grant numbers: 90606002, 11075076, and 10775070) and the National Key Projects for Basic Research of China (Grant numbers: 2009CB930501, 2010CB923401) for supporting this project. The Program for New Century Excellent Talents in University of China Grant No. NCET-07-0422 is also acknowledged.

**References&Notes**

1  H. J. Goldsmid, J. Appl. Phys. **32**, 2198 (1961).

2  H. J. GoldSmid, Proc. Phys. Soc. **72**, 17 (1958).

3  D. Teweldebrhan, V. Goyal, and A. A. Balandin, Nano Lett **10**, 1209 (2010).

4  V. Goyal, D. Teweldebrhan and A. A. Balandin, Appl. Phys. Lett. **97**, 133117 (2010).

5  J. Dheepa, R. Sathyamoorthy, and A. Subbarayan, J. Crys. Growth **274**, 100 (2005).

6  Y. L. Chen, J. G. Analytis, J. H. Chu, Z. K. Liu, S. K. Mo, X. L. Qi, H. J. Zhang, D. H. Lu, X. Dai, Z. Fang, S. C. Zhang, I. R. Fisher, Z. Hussain, and Z. X. Shen, Science **325**, 178 (2009).

7  D. X. Qu, Y. S. Hor, J. Xiong, R. J. Cava, N. P. Ong, Science **329**, 821 (2010).

8  H. J. Zhang, C. X. Liu, X. L. Qi, X. Dai, Z. Fang, and S. C. Zhang, Nat. Phys. **5**, 439 (2009).

9  M. Z. Hasan, C. L. Kane, Rev. Mod. Phys. **82**, 3045 (2010).

10  X. L. Qi, S. C. Zhang, Physics Today p33, Jan. 2010.

11  J. Moore, Nature **464**, 194 (2010).

12  D. Hsieh, Y. Xia, D. Qian, L. Wray, J. H. Dil, F. Meier, J. Osterwalder, L. Patthey, J. G. Checkelsky, N. P. Ong, A. V. Fedorov, H. Lin, A. Bansil, D. Grauer, Y. S. Hor, R. J. Cava, and M. Z. Hasan, Nature **460**, 1101 (2009).

13  Y. Xia, D. Qian, D. Hsieh, L. Wray, A. Pal, H. Lin, A. Bansil, D. Grauer, Y. S. Hor, R. J. Cava, and M. Z. Hasan, Nat Phys **5**, 398 (2009).

14   Y. Y. Li, G. Wang, X. G. Zhu, M. H. Liu, C. Ye, X. Chen, Y. Y. Wang, K. He, L. L. Wang, X. C. Ma, H. J. Zhang, X. Dai, Z. Fang, X. C. Xie, Y. Liu, X. L. Qi, J. F. Jia, S. C. Zhang, Q. K. Xue,   Adv. Mater. **22**, 4002 (2010).

15   D. Teweldebrhan, V. Goyal, M. Rahman, and A. A. Balandina,   Appl. Phys. Lett. **96**, 053107 (2010).

16   A. K. Geim and K. S. Novoselov,   Nat. Mater. **6**, 183 (2007).

17   R. K. Nkum, A. A. Adimado, and H. Totoe,   Mater. Sci. Eng.B **55**, 102 (1998).

18   Y. B. Xu, Z. M. Ren, W. L. Ren, G. H. Cao, K. Deng, and Y. B. Zhong,   Mater Lett **62**, 4273 (2008).

19   F. Q. Song, Powles. R., X. F. Wang, N. A. Marks, L. B. He, S. F. Zhao, J. F. Zhou, Wan. J. G., S. P. Ringer, M. Han, and Wang. G. H., Appl. Phys. Lett. **96**, 033103 (2010).

20   E. Stolyarova, K. T. Rim, S. M. Ryu, J. Maultzsch, P. Kim, L. E. Brus, T. F. Heinz, M. S. Hybertsen, and G. W. Flynn,   PNAS **104**, 9209 (2007).

21   P. Blake, E. W. Hill, A. H. C. Neto, K. S. Novoselov, D. Jiang, R. Yang, T. J. Booth, and A. K. Geim,   Appl. Phys. Lett. **91**, 063124 (2007).

22   Z. H. Ni, H. M. Wang, J. Kasim, H. M. Fan, T. Yu, Y. H. Wu, Y. P. Feng, and Z. X. Shen,   Nano Letters **7**, 2758 (2007).

23   K. M. F. Shahil, M. Z. Hossain, D. Teweldebrhan, A. A. Balandin,   Appl. Phys. Lett. **96**, 153103 (2010).

24   D. S. L. Abergel, A. Russell, and V. I. Fal'ko,   Appl. Phys. Lett. **91**, 063125 (2007).

25  E. D. Palik, *Handbook of Optical Constants of Solids*, Academic Press, New York (1991).

26  E. H. Kaddouri, T. Maurice, X. Gratens, S. Charar, S. Benet, A. Mefleh, J. C. Tedenac, and B. Liautard, Phys. Stat. Sol. A **176**, 1071 (1999).

27  M. C. Payne, M. P. Teter, D. C. Allan, T. A. Arias, and J. D. Joannopoulos, Rev. Mod. Phys. **64**, 1045 (1992).

**Figure Caption**

**Figure 1**. A few-quintuple-layer $Bi_2Te_3$ flake. (a) is its AFM image and (b) is its optical image. The inset in, (a) is its profile which correspond to the blue line. The scale bars of (a) and (b) are 2μm. The edge of the flake is not sharp due to the fact that the size of the flake is close to the diffraction limit.

**Figure 2**. The calculated optical contrast. (a). The calculated three-dimensional diagram of the optical contrast of the $Bi_2Te_3$ film against the incident wavelength and the membrane thickness (marked as height) when the cap silica is 300nm thick. The inset left show the relation of optical contrast against the layer number with 570nm monochromatic light incidence. The inset right is the experimental configuration. (b). The calculated optical contrast of a quintuple layer is plotted against the silica thickness and the incident wavelength.

**Figure 3**. Optical micrographs of a wafer with the TI flakes upon the illumination by the white light filtered with the center wavelength of 570nm (a), 524nm (b) and 470nm(c). The white arrows mark the same $Bi_2Te_3$ flake for comparison. All the bars are 20μm.

**Figure 4**. The contrast reversed with the thickness increase. (a). Optical micrograph of a wafer with many $Bi_2Te_3$ flakes illuminated by a low-power blue laser. One can see the reverse contrasts of some flakes marked by the arrows. The top one measures 20nm thick and the bottom one measures 5nm thick. The width of the micrograph is around 70μm. (b). The experiment contrast against the quintuple layers (marked as thickness) with a λ=473nm laser incident. The blue blocks, red-dot line and black line correspond to the experiment data, first-principle calculation and semi-empirical calculation, respectively.

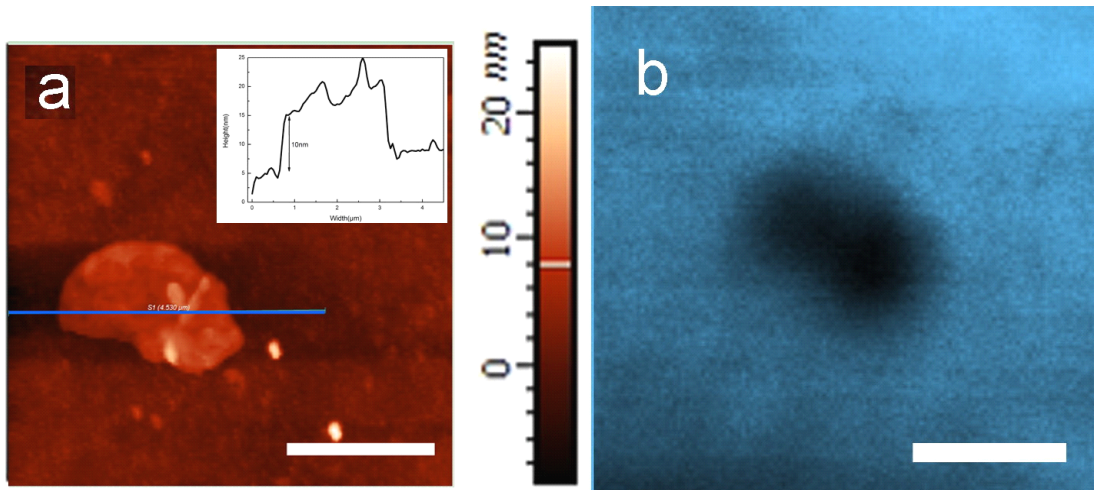

Li et al Fig 1

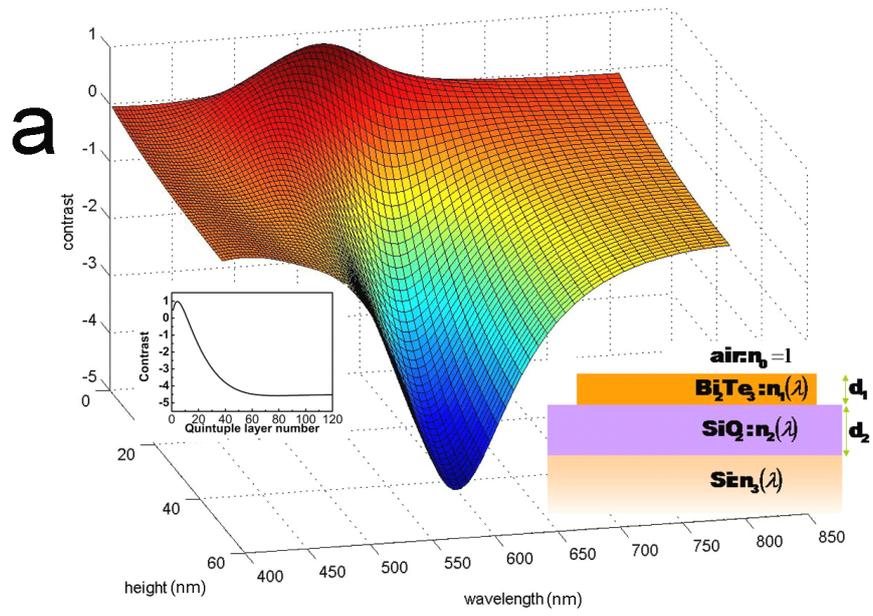

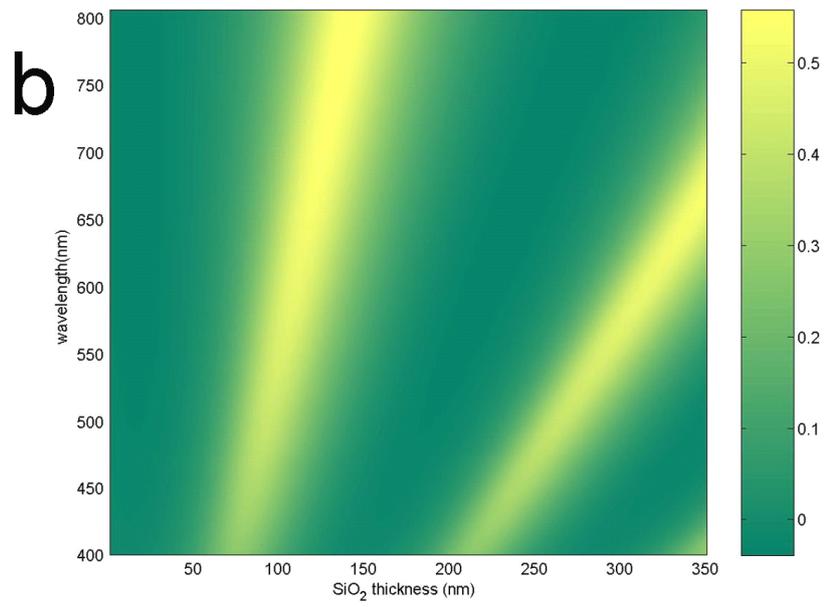

**Li et al , Fig 2**

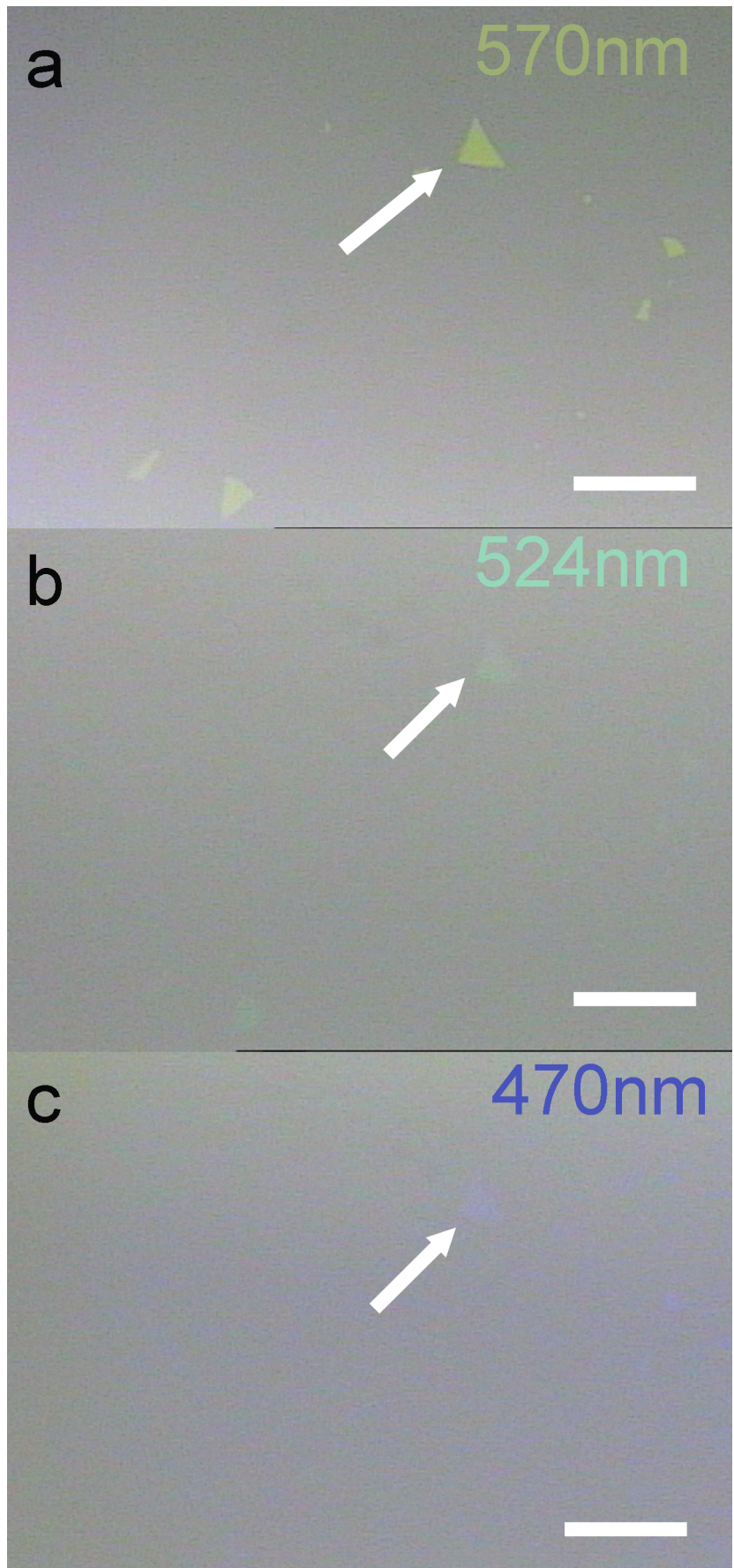

Li et al Fig 3

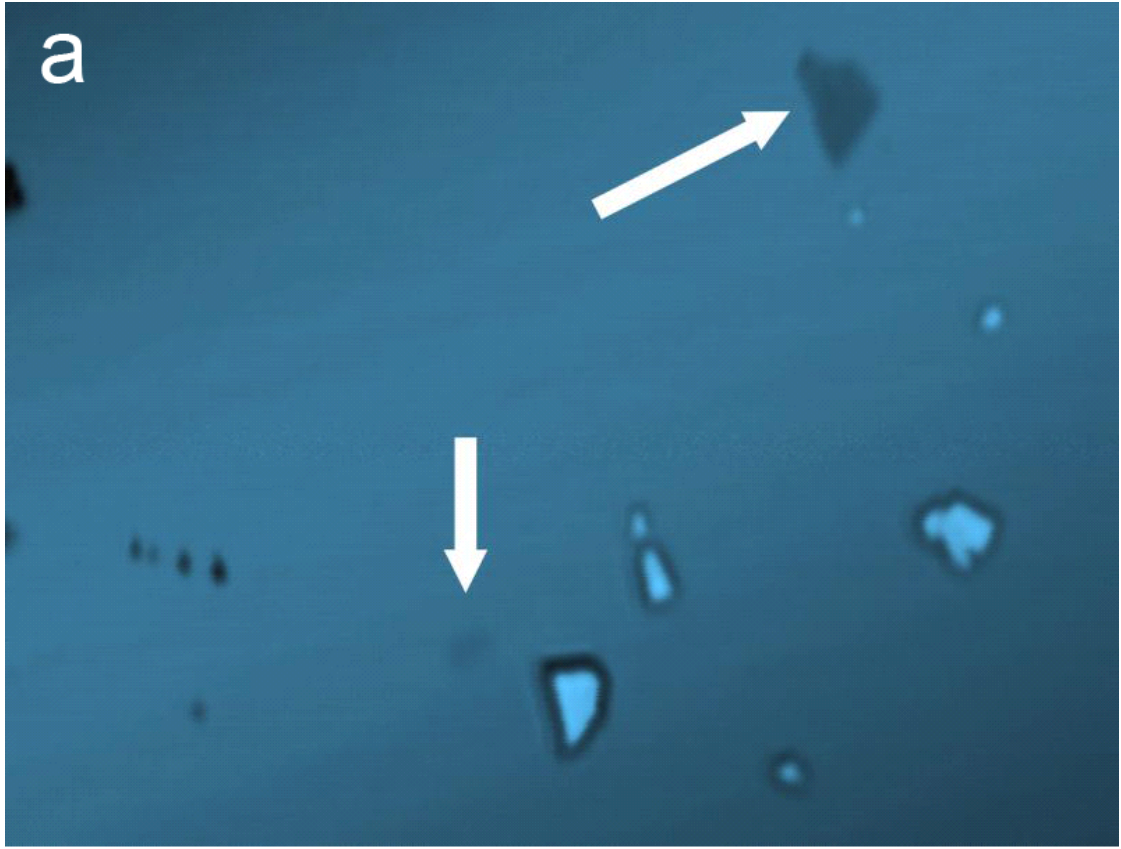

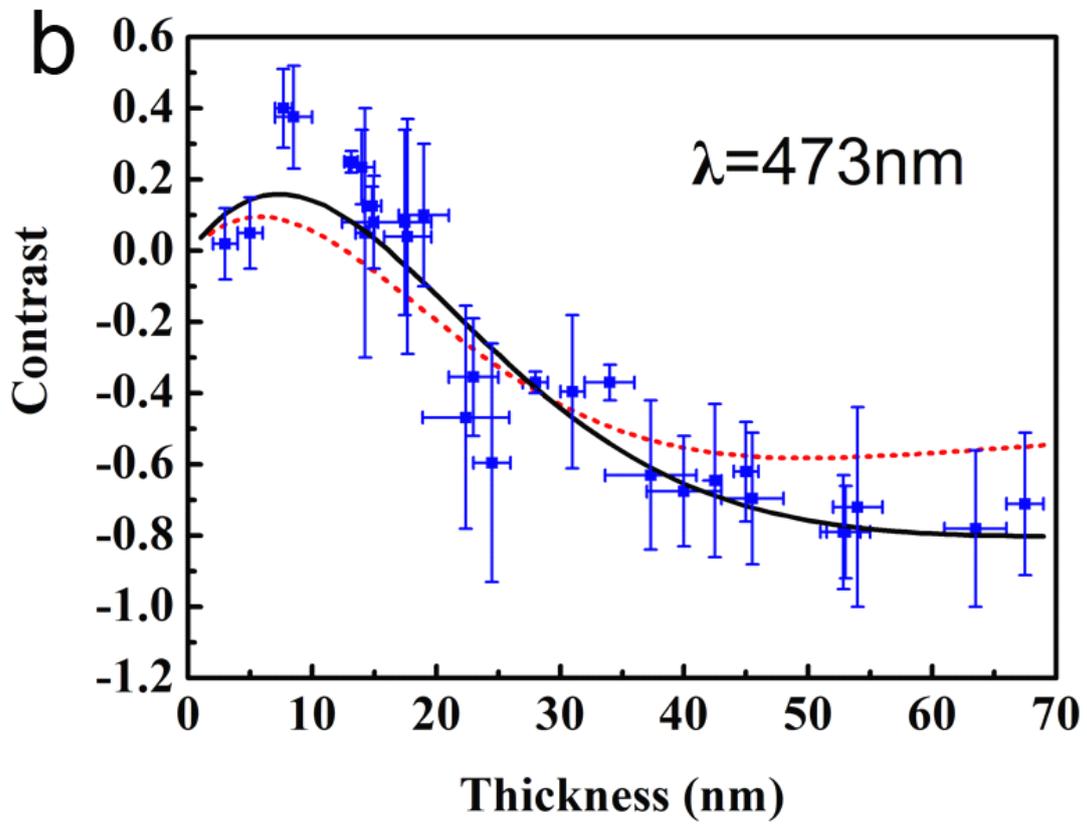

**Li et al , Fig 4**